\documentclass[prb,twocolumn,superscriptaddress,showpacs,floatfix]{revtex4}
\usepackage{graphics}
\usepackage{graphicx}

\begin{document}

\title {Heterointerface effects on the charging energy
of shallow D$^-$ ground state in silicon: the role of
dielectric mismatch}

\author{M.J. Calder\'on}
\affiliation{Instituto de Ciencia de Materiales de Madrid (CSIC), Cantoblanco, 28049 Madrid, Spain}
\author{J. Verduijn}
\affiliation{Kavli Institute of Nanoscience, Delft University of Technology, Lorentzweg 1, 2628 CJ Delft, The Netherlands}
\author{G.P. Lansbergen}
\affiliation{Kavli Institute of Nanoscience, Delft University of Technology, Lorentzweg 1, 2628 CJ Delft, The Netherlands}
\author{G.C. Tettamanzi}
\affiliation{Kavli Institute of Nanoscience, Delft University of Technology, Lorentzweg 1, 2628 CJ Delft, The Netherlands}
\author{S. Rogge}
\affiliation{Kavli Institute of Nanoscience, Delft University of Technology, Lorentzweg 1, 2628 CJ Delft, The Netherlands}
\author{Belita Koiller}
\affiliation{Instituto de F\'{\i}sica, Universidade Federal do Rio de
Janeiro, Caixa Postal 68528, 21941-972 Rio de Janeiro, Brazil}
\date{\today}

\begin{abstract}
Donor states in Si  nanodevices can be strongly modified by
nearby insulating barriers and metallic gates. We report here
experimental results indicating a strong reduction in the
charging energy of isolated As dopants in Si FinFETs, relative
to the bulk value. By studying the problem of two electrons
bound to a shallow donor within the effective mass approach, we
find that the measured small charging energy may be due to a
combined effect of the insulator screening and the proximity of
metallic gates.
\end{abstract}


\maketitle

\section{Introduction}
\label{sec:introduction}
For over a decade dopants in Si have constituted the key elements in
proposals for the implementation of a solid state quantum
computer.~\cite{kane98,vrijen00,skinner03,barrett03,hollenberg041} Spin or charge qubits operate through controlled
manipulation (by applied electric and magnetic fields) of the
donor electron bound states. A shallow donor, as P or As in Si,
can bind one electron in the neutral state, denoted by $D^0$,
or two electrons in the negatively charged state, denoted by
$D^-$. Proposed one and two-qubit gates involve manipulating
individual electrons or electron pairs bound to donors or drawn
away towards the interface of Si with a barrier
material.~\cite{kane98,skinner03,calderonPRL06} In general, neutral and ionized donor
states play a role in different stages of the prescribed
sequence of operations.

In the proposed quantum computing schemes, donors are located
very close to interfaces with insulators, separating the Si
layer from the control metallic gates. This proximity is
required in order to perform the manipulation via electric
fields of the donor spin and charge states. The presence of
boundaries close to donors modifies the binding potential
experienced by the electrons in a semiconductor. This is a
well-known effect in Si
MOSFETs,~\cite{macmillen84,calderon-longPRB07} where the
binding energy of electrons is reduced with respect to the bulk
value for distances between the donor and the interface smaller
than the typical Bohr radius of the bound electron
wave-function. On the other hand, in free-standing Si nanowires
with diameters below $10$ nm, the binding energy of donor
electrons significantly increases~\cite{delerue-lannoo,diarra07} leading to a
strongly reduced doping efficiency in the nanowires.~\cite{bjork2009}

The continuous size reduction of transistors along years, with
current characteristic channel lengths of tens of nanometers,
implies that the disorder in the distribution of dopants can
now determine the performance, in particular the transport
properties of the devices.~\cite{voyles02,shinada05,pierre2009}
In specific geometries, like the nonplanar field effect
transistors denoted by FinFETs,~\cite{sellier2006} isolated
donors can be identified and its charge states (neutral $D^0$, and negatively charged $D^-$)
studied by transport spectroscopy.

The existence of $D^-$ donor states in semiconductors,
analogous to the hydrogen negative ion $H^-$, was suggested in
the fifties~\cite{lampertPRL58} and is now well established
experimentally. Negatively charged donors in bulk Si were first
detected by photoconductivity
measurements.~\cite{taniguchi-narita76} The binding energies of
$D^-$ donors, defined as the energy required to remove one
electron from the ion ($D^- \rightarrow D^0+$ free-electron)
$E_B^{D^-}=E_{D^0}-E_{D^-}$, are found experimentally to be
small ($E_B^{D^-}\sim 1.7$ meV for P and $\sim 2.05$ meV for
As) compared to the binding energies of the first electron
$E_B^{D^0}$ (45 meV for P and 54 meV for As). For zero applied
magnetic fields, no excited bound states of $D^-$ in
bulk semiconductors~\cite{larsenPRB92} or superlattices~\cite{peetersPRB95} are found, similar to
$H^-$ which has only one bound state in three-dimensions as shown in
Refs.~\onlinecite{perkerisPR62,hillPRL77}.

A relevant characteristic of negatively charged donors is their charging
energy,  $U= E_{D^-}-2 E_{D^0}$,  which gives the energy
required to add a second electron to a neutral donor. This
extra energy is due to the Coulomb repulsion between the two
bound electrons, and does not contribute in one electron systems,
as $D^0$. The measured values in bulk Si are $U_{\rm As}^{\rm
bulk, exp} = 52$ meV for As and $U_{\rm P}^{\rm bulk, exp} =
43$ meV for P.

From the stability diagrams obtained from transport
spectroscopy measurements we observe that the charging energy
of As dopants in nanoscale Si devices (FinFETs) is strongly
reduced compared to the well known bulk value. By using a
variational approach within the single-valley effective mass
approximation, we find that this decrease of the charging
energy may be attributed to modifications on the bare insulator
screening by the presence of a nearby metallic layer. For the
same reason, we also find theoretically that it may be possible
to have a $D^-$ bound excited state.

This paper is organized as follows. In Sec.~\ref{sec:bulk}, we
introduce the formalism for a donor in the bulk in analogy with
the hydrogen atom problem. In Sec.~\ref{sec:interface}, we
study the problem of a donor close to an interface within a flat band condition. We show
experimental results for the charging energy and compare them
with our theoretical estimations. We also calculate the binding energy of a $D^-$ triplet first excited state. In Sec.~\ref{sec:discussion}
we present discussions including: (i) assessment of the limitations in our theoretical approach, (ii) considerations about the
modifications of the screening in nanoscale devices, (iii) the implications of our results in
quantum device applications, and, finally, we also
summarize our main conclusions.

\section{Donors in bulk silicon}
\label{sec:bulk}

A simple estimate for the binding energies of both $D^0$ and
$D^-$ in bulk Si can be obtained using the analogy between the
hydrogen atom $H$ and  shallow donor states in semiconductors.
The Hamiltonian for one electron in the field of a nucleus with
charge $+e$ and infinite mass is, in effective units of length
$a_B=\hbar^2/m_e e^2$ and energy $Ry=m_e e^4/2\hbar^2$,
\begin{equation}
h(r_1)=T(r_1)-\frac{2}{r_1} \, ,
\label{eq:hamil-1e}
\end{equation}
with $T(r)=- \nabla^2 $. The ground state is
\begin{equation}
\phi(r_1,a)=\frac{1}{\sqrt{\pi a^3}} e^{-r_1/a}
\label{eq:wfH}
\end{equation}
with Bohr radius $a=1\,a_B$ and energy $E_{H}=-1\, Ry$. This
corresponds to one electron in the $1s$ orbital.

For negatively charged hydrogen (H$^-$) the two electrons Hamiltonian is
\begin{equation}
H_{\rm Bulk}=h(r_1)+h(r_2)+\frac{2}{r_{12}} \,,
\end{equation}
where the last term gives the electron-electron interaction
($r_{12}=|\vec{r}_1-\vec{r}_2|$). As an approximation to the
ground state, we use a relatively
simple variational two particles wave-function for the spatial part, a symmetrized combination of $1s$ atomic orbitals as given in Eq.~\ref{eq:wfH}, since the spin
part is a singlet,
\begin{equation}
|1s,1s,s\rangle =\left[\phi(r_1,a) \phi(r_2,b) + \phi(r_1,b) \phi(r_2,a) \right] \,.
\label{eq:wfH-}
\end{equation}

The resulting energy is $E^{H^-}=-1.027 Ry$ with $a=0.963\,
a_B$ and $b=3.534 \,a_B$ (binding energy $E_B^{H^-}=0.027
Ry$).~\cite{Bethe-Salpeter} Here we may interpret $a$ as the
radius of the inner orbital and $b$ of the outer orbital. This
approximation for the wave-function correctly gives a bound
state for $H^-$ but it underestimates the binding energy with
respect to the value $E_B^{H^-}=0.0555$ Ry, obtained with
variational wave-functions with a larger number of parameters,
thus closer to the 'exact' value.~\cite{Bethe-Salpeter}

Assuming an isotropic single-valley conduction band in bulk Si
the calculation of the $D^0$ and $D^-$ energies reduces to the
case of $H$ just described. Within this approximation, an
estimation for $E_B^{D^-}$ can then be obtained by considering
an effective rydberg $Ry^*=m^*e^4/2\epsilon_{\rm Si}^2 \hbar^2$
with an isotropic effective mass (we use $\epsilon_{\rm
Si}=11.4$). We choose $m^* = 0.29819 \,m_e$ so that the ground
state energy for a neutral donor is the same as given by an
anisotropic wave-function in bulk: within a single valley
approximation $E_B^{D^0} = -1Ry^* = -31.2$ meV and its
effective Bohr radius is $a=1 a^*$ with
$a^*={{\hbar^2\epsilon_{\rm Si}}/{m^* e^2}} =2.14$ nm. In this
approximation, $E_B^{D^-}=0.84 $ meV. In the same way, an
estimation for the charging energy can be made for donors in
Si: $U=0.973 Ry^*= 30.35$ meV.~\cite{ramdas1981}

Even though the trial wave function in Eq.~(\ref{eq:wfH-})
underestimates the binding energy, we adopt it here for
simplicity, in particular to allow performing in a reasonably
simple way the calculations for a negatively charged donor
close to an interface reported below. In the same way, we do
not introduce the multivalley structure of the conduction band
of Si. The approximations proposed here lead to qualitative
estimates and establish general trends for the effects of an
interface on a donor energy spectrum. The limitations and
consequences of our approach are discussed in
Sec.~\ref{sec:discussion}.
\begin{table}[t]
\begin{tabular}{|c l l l|}
\hline
E$_{\rm{D}^0}$ & $=-1 Ry^*$  & &  $a=1 a^*$\\
\hline
E$_{\rm{D}^-}$ & $=-1.027 Ry^*$ & & $a=0.963 a^*$; $b=3.534 a^*$ \\
\hline
E$_B$ & = E$_{\rm{D}^0}$- E$_{\rm{D}^-}$ &$ =0.027 Ry^*$ & \\
\hline
U   & =  E$_{\rm{D}^-}$-2E$_{\rm{D}^0}$ & $ =0.973 Ry^*$  & \\
\hline
\end{tabular}\caption{Bulk values of energies and orbital radii for the ground
state of neutral and negatively charged donors within our
approximation (see text for discussion). Effective units for
Si are $a^*=2.14$ nm and $Ry^*=31.2$ meV. } \label{table:data}
\end{table}
\begin{figure}
\resizebox{60mm}{!}{\includegraphics{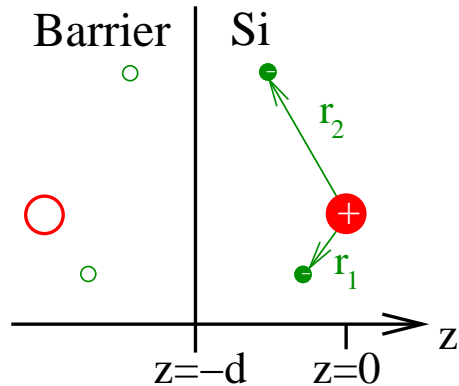}}
\caption{(Color online)
Schematic representation of a negatively charged donor in Si
(solid circles) located a distance $d$ from an interface. The open circles in the
barrier (left) represent the image charges. The sign and
magnitude of these charges depend on the relation between the
dielectric constants of Si and the barrier given by
$Q=(\epsilon_{\rm barrier}-\epsilon_{\rm Si})/(\epsilon_{\rm
barrier}+\epsilon_{\rm Si})$.  For the electrons, $Q<0$ corresponds to repulsive electron image potentials and a positive donor image potential (opposite signs of potentials and image charges for $Q>0$, see Eq.~\ref{eq:beforeQ}).
}
\label{fig:scheme}
\end{figure}
\section{Donors close to an interface}
\label{sec:interface}
\subsection{$D^0$ and $D^-$ ground states}
We consider now a donor (at $z=0$) close to an interface (at
$z=-d$) (see Fig.~\ref{fig:scheme}). Assuming that the interface produces an
infinite barrier potential, we adopt variational wave-functions
with the same form as in Eqs.~(\ref{eq:wfH}) and
(\ref{eq:wfH-}) multiplied by linear factors $(z_i+d)$
($i=1,2$) which guarantee that each orbital goes to zero at the
interface. We further characterize the Si interface with a
different material by including charge image terms in the
Hamiltonian.
\begin{figure}
\resizebox{80mm}{!}{\includegraphics{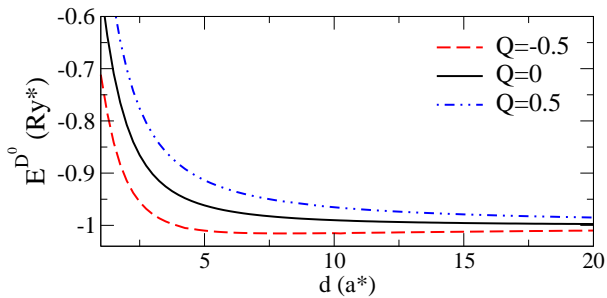}}
\caption{(Color online) Energy of the neutral donor versus its distance $d$ from an interface for different values of $Q=(\epsilon_{\rm barrier}-\epsilon_{Si})/(\epsilon_{\rm barrier}+\epsilon_{Si})$. }
\label{fig:E-D0}
\end{figure}

Before discussing the ionized donor $D^-$, we briefly present
results for the neutral donor $D^0$ which are involved in
defining donor binding and charging energies. For this case,
the Hamiltonian is
\begin{equation}
H(r_1)=h(r_1)+h_{\rm images} (r_1)
\label{eq:hamil+images-1e}
\end{equation}
with $h(r_1)$ as in Eq.~\ref{eq:hamil-1e} and
\begin{equation}
h_{\rm images}(r_1)=-\frac{Q}{2(z_1+d)}+\frac{2 Q}{\sqrt{x_1^2+y_1^2+(z_1+2d)^2}} \, ,
\label{eq:beforeQ}
\end{equation}
where $Q=(\epsilon_{\rm barrier}-\epsilon_{Si})/(\epsilon_{\rm
barrier}+\epsilon_{Si})$. $\epsilon_{\rm barrier}$ is the
dielectric constant of the barrier material. The first term in
$h_{\rm images}$ is the interaction of the electron with its
own image, and the second is the interaction of the electron
with the donor's image. If the barrier is a thick insulator,
for example SiO$_2$ with dielectric constant $\epsilon_{\rm
SiO_2}=3.8$, $Q<0$ ($Q= -0.5$ in this case). In actual
devices, the barrier is composed of a thin insulator (usually
SiO$_2$), which prevents charge leakage, plus metallic
electrodes which control transport and charge in the
semiconductor. This composite heterostructure may effectively
behave as a barrier with an effective dielectric constant
larger than Si, since $\epsilon_{\rm metal} \to \infty$,
leading to an effective $Q>0$. Depending on the sign of $Q$,
the net image potentials will be repulsive or attractive, which
may strongly affect the binding energies of donors at a short
distance $d$ from the interface.

 Using a trial wave-function $\phi_{D^0} \propto e^{-r/a}
(z+d)$, most of the integrals involved in the variational
calculation of $E^{D^0}$ can be performed analytically.
$E^{D^0}$ is shown in Fig.~\ref{fig:E-D0} for different values
of $Q$ and compare very well with the energy calculated  by
MacMillen and Landman~\cite{macmillen84} for $Q= -1$ with a
much more complex trial wave-function. The main effect of the
interface is to reduce the binding energy when the donor is
located at very small distances $d$. For $Q<0$ (corresponding
to insulating barriers with a dielectric constant smaller than
that of Si), the energy has a shallow minimum for $d\sim 8
a^*$. This minimum arises because the donor image attractive
potential enhances the binding energy but, as $d$ gets smaller,
the fact that the electron's wave-function is constrained to
$z>-d$ dominates, leading to a strong decrease in the binding
energy.~\cite{macmillen84}  $Q=0$ corresponds to ignoring the
images. $Q=1$ would correspond to having a metal at the
interface with an infinitesimal insulating barrier at the
interface to prevent leackage of the wave-function into the
metal.~\cite{slachmuylders08} We show results for $Q=0.5$ as an
effective value to take account of a realistic barrier composed
of a thin (but finite) insulator plus a metal. The bulk limit
$E=-1 \, Ry^*$ is reached at long distances for all values of
$Q$.

Adding a second electron to a donor requires the inclusion of
the electron-electron interaction terms.
The negative donor Hamiltonian parameters are schematically presented in
Fig.~\ref{fig:scheme}, and the total two electrons Hamiltonian is
\begin{eqnarray}
H&=&H(r_1)+H(r_2)+\frac{2}{r_{12}}\nonumber \\
&-&\frac{4 Q}{\sqrt{(x_1-x_2)^2+(y_1-y_2)^2+(2d+z_1+z_2)^2}}\, ,
\end{eqnarray}
where $H(r_i)$ includes the one-particle images (Eq.~\ref{eq:hamil+images-1e}) and the last term is the interaction between each electron and the other electron's image.

\begin{figure}
%\resizebox{90mm}{!}{\includegraphics{E-E_B-a-b-Q-05.eps}}
\resizebox{90mm}{!}{\includegraphics{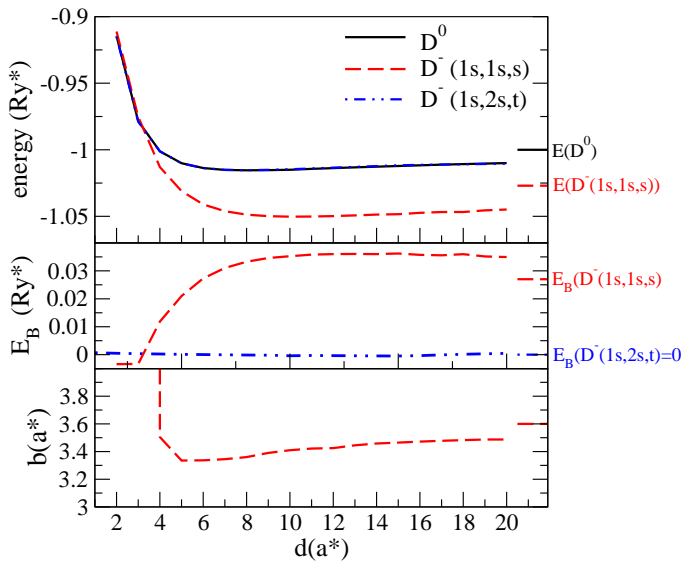}}
\caption{(Color online) Results for $Q=-0.5$.
(a) Energy for a neutral donor $D^0$, and for the ground $D^-|1s,1s,s\rangle$
and first excited $D^{-}|1s,2s,t\rangle$ negatively charged donor. (b) Binding
energies of the $D^-$ states. (c) Value of the variational parameter $b$ for the
$D^-$ ground state. For $d <4$, $D^- |1s,1s,s\rangle$ is not stable and the energy
is minimized with $b \rightarrow \infty$. Bulk values are represented by short
line segments on the right.}
\label{fig:E-EB-b-Q-05}
\end{figure}

\begin{figure}
%\resizebox{90mm}{!}{\includegraphics{E-E_B-a-b-Q05.eps}}
\resizebox{90mm}{!}{\includegraphics{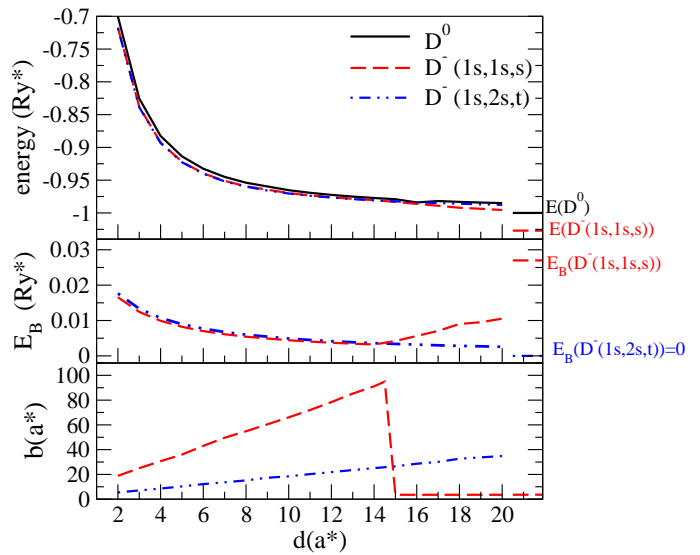}}
\caption{(Color online) Same as Fig.~\ref{fig:E-EB-b-Q-05} for $Q=0.5$. }
\label{fig:E-EB-b-Q05}
\end{figure}

In Figs.~\ref{fig:E-EB-b-Q-05} and \ref{fig:E-EB-b-Q05}, we
plot $E^{D^-}$ and the binding energy
$E_B^{D^-}=E^{D^0}-E^{D^-}$ assuming a trial wave-function
$\propto \left[\phi(r_1,a) \phi(r_2,b) + \phi(r_1,b)
\phi(r_2,a) \right ] (z_1+d) (z_2+d)$ with variational
parameters $a$ and $b$, for $Q=-0.5$ and $Q=0.5$ respectively.
The radius of the  inner orbital is $a \sim 1a^*$  while $b$,
the radius of the  outer orbital, depends very strongly on $Q$
and $d$ and is shown in Figs.~\ref{fig:E-EB-b-Q-05}(c) and
\ref{fig:E-EB-b-Q05}(c). We have done calculations for several
values of $Q$, ranging from $Q=+1$ to $Q=-1$. The general
trends and qualitative behavior of the calculated quantities
versus distance $d$ are the same for all $Q > 0$ (effective
barrier dominated by the metallic character of the interface
materials), which differ from the also general behavior of $Q
\leq 0$ (effective barrier dominated by the insulator
material). For $Q \leq 0$ (illustrated for the particular case
of $Q=-0.5$ in Fig.~\ref{fig:E-EB-b-Q-05}), $D^-$ is not bound
for small $d$ (for $d <4\,a^*$ in the case of $Q=-0.5$). For
larger $d$'s, the binding energy is slightly enhanced from the
bulk value. The radius of the outer orbital $b$ is very close
to the bulk value for $d \geq 4\,a^*$.  For $Q>0$ (illustrated
by $Q=0.5$ in  Fig.~\ref{fig:E-EB-b-Q05}), $D^-$ is bound at
all distances $d$, though the binding energy is smaller than in
bulk. The radius of the outer orbital $b$ is very large and
increases linearly with $d$ up to $d_{\rm crossover} \sim 14.5
a^*$ [see Fig.~\ref{fig:E-EB-b-Q05} (c)]. For larger $d$, $b$ is suddenly reduced to its bulk
value. This abrupt behavior of the $b$ that minimizes the
energy is due to two local minima in the energy versus $b$: for
$d < d_{\rm crossover}$ the absolute minimum corresponds to a
very large (but finite) orbital radius $b$ while for $d >
d_{\rm crossover}$ the absolute minimum crosses over to the
other local minimum, at $b\sim b_{\rm bulk}$. As $d$ increases
from the smallest values and $b$ increases up to the
discontinuous drop, a ``kink'' in the D$^-$ binding energy  is
obtained at the  crossover point [see Fig.~\ref{fig:E-EB-b-Q05} (b)], changing its behavior from a
decreasing to an increasing dependence on $d$ towards the bulk
value as $d\to\infty$.

\subsection{Charging energy: experimental results}
\label{sec:experiments}

The charging energy of shallow dopants can be obtained by using
the combined results of photoconductivity experiments to
determine the $D^-$ binding energy~\cite{dean1967} and direct
optical spectroscopy to determine the binding of the $D^0$
state.~\cite{ramdas1981} It was shown recently that the
charging energy in nanostructures can be obtained directly from
charge transport spectroscopy at low
temperature.~\cite{sellier2006} Single dopants can be accessed
electronically at low temperature in deterministically doped
silicon/silicon-dioxide heterostructures~\cite{morello10} and
in small silicon nanowire field effect transistors (FinFETs),
where the dopants are positioned randomly in the
channel.~\cite{sellier2006,lansbergen-NatPhys,pierre2009} Here
we will focus in particular on data obtained using the latter
structures.~\cite{sellier2006,lansbergen-NatPhys}

FinFET devices in which single dopant transport have been
observed typically consist of crystalline silicon wire channels
with large patterned contacts fabricated on
silicon-on-insulator. Details of the fabrication can be found
in Ref.~\onlinecite{sellier2006}.
In this kind of samples, few dopants may diffuse from the
source/drain contacts into the channel during the fabrication
modifing the device characteristics both at
room~\cite{pierre2009} and low
temperatures.~\cite{sellier2006,lansbergen-NatPhys,pierre2009}
In some cases, subthreshold transport is dominated by a single
dopant.~\cite{lansbergen-NatPhys}

Low temperature transport spectroscopy relies on the presence
of efficient Coulomb blockade with approximately zero current
in the blocked region. This requires the thermal energy of the
electrons, $k_BT$, to be much smaller than $U$, a requirement
that is typically satisfied for shallow dopants in silicon at
liquid helium temperature and below, i.e. $\leq 4.2K$. At these
temperatures the current is blocked in a diamond-shaped region
in a stability diagram, a color-scale plot of
the current -- or differential conductance
$\text{d}I/\text{d}V_b$ -- as a function of the source/drain,
$V_b$, and gate voltage, $V_g$.

\begin{figure}
\resizebox{90mm}{!}{\includegraphics{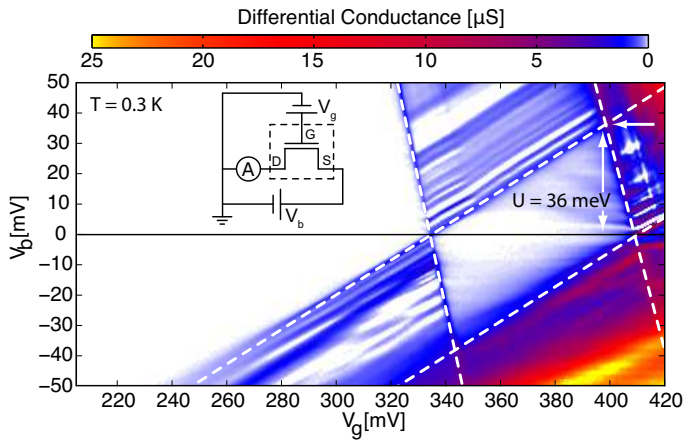}}
\caption{(Color online) Differential conductance stability diagram showing the transport characteristics of a single As donor in a FinFET device.~\cite{sellier2006} The differential conductance is obtained by a numerical differentiation of the current with respect to $V_b$ at a temperature of $0.3$ K. Extracting the charging energy from the stability diagram can be done by determining the gate voltage for which Coulomb blockade of a given charge state (the $D^0$ charge state in this case) is lifted for all $V_g$. The transition point is indicated by the horizontal arrow,
leading to a charging energy $U=36$ meV, as given by the vertical
double-arrow. The inset shows the electrical circuit used for the measurement.}
\label{fig:experimentaldata}
\end{figure}

In Fig.~\ref{fig:experimentaldata}, the stability diagram of a
FinFET with only one As dopant in the conduction channel is
shown. At small bias voltage ($eV_b \ll k_BT$), increasing the
voltage on the gate effectively lowers the potential of the
donor such that the different donor charge states can become
degenerate with respect to the chemical potentials in the
source and drain contacts and current can flow. The difference
in gate voltage between the D$^+/$D$^0$ and D$^0/$D$^-$
degeneracy points (related to the charging energy) depends,
usually in good approximation, linearly on the gate voltage
times a constant capacitive coupling to the
donor.~\cite{sellier2006} Generally a more accurate and direct
way to determine the charging energy is to determine the bias
voltage at which the Coulomb blockade for a given charge state
is lifted for all gate voltages, indicated by the horizontal arrow in
Fig.~\ref{fig:experimentaldata}. This method is especially
useful when there is efficient Coulomb blockade. For the
particular sample shown in Fig.~\ref{fig:experimentaldata},
$U=36$ meV. This is similar to other reported values in the
literature~\cite{sellier2006,lansbergen-NatPhys,pierre2009}
ranging from $\sim 26$ to $\sim 36$ meV. There is therefore a
strong reduction in the charging energy compared to the bulk
value $U_{\rm bulk} =52$ meV. The ratio between the observed
and the bulk value is $\sim 0.6-0.7$.

\begin{figure}
\resizebox{90mm}{!}{\includegraphics{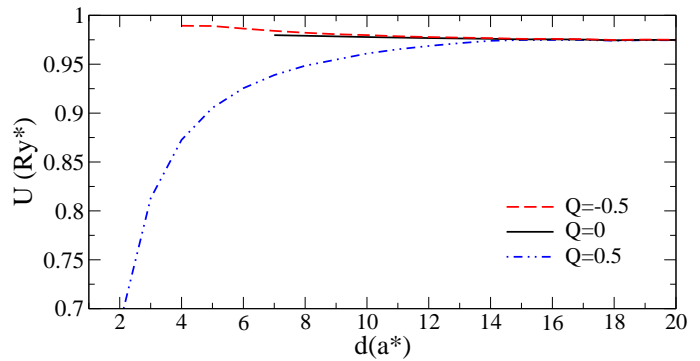}}
\caption{(Color online)
Charging energy $U$ of the D$^-$ ground state for three different values of $Q$. For $Q
\leq 0$, the charging energy is nearly constant with $d$. For
these cases, the negatively charged donor is not bound for
small $d$. For $Q > 0$ the charging energy decreases as  the donor gets closer to the interface,
at relatively small distances $d$. The latter is consistent with the
experimental observation.}\label{fig:U}
\end{figure}

Theoretically, we can extract the charging energy from the
results in Figs.~\ref{fig:E-EB-b-Q-05} and
\ref{fig:E-EB-b-Q05}. The results are shown as a function of
$d$ for $Q=-0.5$, $0$, and $0.5$ in Fig.~\ref{fig:U}. A
reduction of the charging energy $U$ of the order of the one
observed occurs at $d \sim 2 a^*$ for $0.1<Q<1$ (only $Q=0.5$
is shown in the figure). Therefore, the experimentally observed
behavior of $U$ is consistent with a predominant influence of
the metallic gates material in the D$^-$ energetics. On the
other hand, for $Q \leq 0$, $U$ is slightly enhanced as $d$
decreases and, for the smallest values of $d$ considered, the
outer orbital is not bound. At very short distances $d$, the
difference in behavior between the insulating barrier ($Q<0$)
and the barrier with more metallic character ($Q>0$) is in the
interaction between each electron and the other electron's
image, which is repulsive in the former case and attractive in
the latter. Although this interaction is small, it is critical
to lead to a bound $D^-$ for $Q>0$ and an unbound $D^-$ for
$Q<0$ at very short $d$.

\subsection{$D^-$ first excited state.}

It is well established that in 3 dimensions (with no magnetic
field applied) there is only one bound state of
$D^-$.\cite{hillPRL77,larsenPRB92}  Motivated by the
significant changes in the ground state energy produced by
nearby interfaces, we explore the possibility of having a bound
excited state in a double-charged single donor. Like helium, we
expect the $D^-$ first excited state to consist of promoting one
 $1s$ electron to the $2s$ orbital. The spin triplet
$|1s,2s,t \rangle$ state (which is orthogonal to the singlet
ground state) has a lower energy than
$|1s,2s,s\rangle$.~\cite{Bransden-Joachain} As a trial
wave-function for $|1s,2s,t \rangle$ we use the antisymmetrized
product of the two orbitals $1s$ and $2s$ and multiply by
$(z_1+d) (z_2+d)$ to fulfill the boundary condition, namely,
\begin{eqnarray}
\Psi_{1s,2s,t}&=& N\left[e^{-\frac{r_1}{a}}
e^{-\frac{r_2}{2b}}\left(\frac{r_2}{2b}-1\right)
- e^{-\frac{r_2}{a}} e^{-\frac{r_1}{2b}}\left(\frac{r_1}{2b}-1\right)\right] \nonumber \\
&\times& (z_1+d) (z_2+d)  \,
\label{eq:wf2Striplet}
\end{eqnarray}
with $a$ and $b$ variational parameters and $N$ a normalization
factor. Note that, for a particular value of $b$, the outer
electron in a $2s$ orbital would have a larger effective orbital radius
than in a $1s$ orbital due to the different form of the radial part.

For $Q < 0$, the outer orbital is not bound  and the energy
reduces to that of $D^0$ (see Fig.~\ref{fig:E-EB-b-Q-05}).
Surprisingly, for $Q > 0$ the $|1s,2s,t \rangle$ state is bound
and, as $d$ increases, tends very slowly to
the $D^0$ energy as shown in Fig.~\ref{fig:E-EB-b-Q05}.
Moreover, its binding energy is roughly the same as the ground
state $|1s,1s,s \rangle$ for $d \leq 15 a^*$, another
unexpected result. The existence of a bound $D^-$ triplet state opens
the possibility of performing coherent rotations involving this state
and the nearby singlet ground state.

\section{Discussions and conclusions}
\label{sec:discussion}

Our model for $D^-$ centers involves a number of simplifications:
(i) the mass anisotropy is not included; (ii) the
multivalley structure of the conduction band of Si is not
considered; (iii) correlation terms in the trial wave-function are neglected. These assumptions aim to decrease the number of variational parameters while allowing many of the integrals to be solved analytically.

Qualitatively, regarding assumption (i), it has been shown that
the mass anisotropy inclusion gives an increase of the binding
energy for both $D^0$ and $D^-$ (see
Ref.~\onlinecite{inouePRB08}); regarding (ii), inclusion of the
multivalley structure of the conduction  together with the
anisotropy of the mass would lead to an enhancement of the
binding energy of $D^-$ due to the possibility of having
intervalley configurations in which the electrons occupy
valleys in 'perpendicular' orientations, (with perpendicularly
oblated wave-functions), thus leading to a strong reduction of
the electron-electron repulsive interaction.
\cite{larsenPRB81,inouePRB08} Regarding point (iii), more general
trial wave-functions for $D^-$ have been proposed in the literature. For example, the one suggested
by Chandrasekar models correlation effects  by multiplying Eq.~\ref{eq:wfH-} by a
 factor,  $(1+Cr_{12})$, ~\cite{chandrasekarRMP44} where $C$ is an additional  variational
parameter. In the bulk, the effect
of this correlation factor is to increase the binding energy of
$D^-$ from $0.027 \, Ry^*$ if $C=0$ (our case) to $0.0518 \,
Ry^*$.~\cite{Bethe-Salpeter} We conclude that all three
simplifications assumed in our model lead to an underestimation
of the binding energy of $D^-$, thus, the values reported here
are to be taken as lower bounds for it.

As compared to experiments, an important difference with
respect to the theory is that we are assuming a flat-band
condition while the actual devices have a built-in electric
field due to band-bending at the interface between the gate
oxide and the p-doped
channel.~\cite{lansbergen-NatPhys,rahman2009} If an electric
field were included, the electron would feel a stronger binding
potential (which results from the addition of the donor
potential and the triangular potential well formed at the
interface) leading to an enhancement of the binding energy of
$D^0$ and $D^-$ (with an expected strong decrease of the
electron-electron interaction in this case for configurations
with one electron bound to the donor at $z=0$ and the other
pulled to the interface at $z=-d$).

The presented results are dominated by the presence of a
barrier, which constrains the electron to the $z>-d$ region,
and the
modification of the screening due to the charge induced at the
interface, a consequence of the dielectric mismatch between Si
and the barrier material. This is included by means of image
charges. Effects of quantum confinement and dielectric
confinement\cite{delerue-lannoo,diarra07} are not considered here: we believe
these are not relevant in the FinFETs under study. Although the
conduction channel is very narrow ($4$ nm$^2$)\cite{sellier07}
the full cross section of the Si wire is various tens of nm
and quantum and dielectric confinement is expected to be
effective for typical device sizes under $10$ nm. Both quantum
and dielectric confinement lead to an enhancement of the
binding energy with respect to the bulk, which is the opposite
to what we obtain for small $d$.

Neutral double donors in Si, such as Te or Se, have been proposed for spin readout via spin-to-charge conversion~\cite{kane00PRB} and for spin coherence time measurements.\cite{calderonDD} 
The negative donor $D^-$ also constitutes a two-electron system, shallower than Te and Se. 
In this context, investigation of the properties of $D^-$ shallow donors in Si affecting quantum operations as, for example, their adequacy for implementing spin measurement via spin-to-charge conversion mechanism,~\cite{kane00PRB,koppens05} deserve special attention.
Our theoretical study indicates that, very near an interface (for $d< 4a^*$), the stability of D$^-$ 
against dissociation requires architectures that yield effective dielectric
mismatch $Q>0$, a requirement for any device involving operations or gates based on D$^-$ bound states.

%\section{Conclusion}
%\label{sec:conclusion}

In conclusion, we have presented a comprehensive study of the effects of
interface dielectric mismatch in the charging energy of nearby
negatively charged donors in Si. In our study, the theoretical
treatment is based on a single-valley effective mass formalism,
while transport spectroscopy experiments were
carried out in FinFET devices. The experiments reveal a strong
reduction on the charging energy of isolated As dopants in
FinFETs as compared to the bulk values. Calculations present,
besides the charging energy, the binding energy of donor in
three different charge states as a function of the distance
between the donor and an interface with a barrier. The boundary
problem is solved by including the charge images whose signs
depend on the difference between the dielectric constant of Si
and that of the barrier material [the dielectric mismatch,
quantified by the parameter $Q$ defined below
Eq.~(\ref{eq:beforeQ})].

Typically, thin insulating layers separate the Si channel,
where the dopants are located, from metallic gates needed to
control the electric fields applied to the device. This
heterostructured barrier leads to an effective screening with
predominance of the metallic components, if compared to a
purely SiO$_2$ thick layer, for which $Q<0$. Assuming a barrier
material with an effective dielectric constant larger than that
of Si (in particular, $Q=0.5$ corresponds to $\epsilon_{\rm barrier}
= 3\epsilon_{\rm Si}$), we obtain a reduction of the charging energy $U$ relative to $U_{\rm
bulk}$ at small $d$, consistent with the experimental
observation. We did not attempt quantitative agreement between
presented values here, but merely to reproduce the right trends
and clarify the underling physics. It is clear from our results
that more elaborate theoretical work on interface effects in
donors, beyond the simplifying assumptions here, should take
into account the effective screening parameter as a combined
effect of the nearby barrier material and the adjacent metallic
electrodes. From our calculations and experimental results, we
conclude that the presence of metallic gates tend to increase
$\epsilon_{\rm barrier}^{\rm effective}$ above $\epsilon_{\rm
Si}$, leading to $Q>0$ and reducing the charging energies.

\acknowledgments
 M.J.C. acknowledges support from Ram\'on y
Cajal Program and FIS2009-08744 (MICINN, Spain). B.K.
acknowledges support from the Brazilian entities CNPq,
Instituto Nacional de Ciencia e Tecnologia em Informa\c c\~ao
Quantica - MCT, and FAPERJ.
J.V., G.P.L, G.C.T and S.R. acknowledge the financial support from the EC FP7 FET-proactive
NanoICT projects MOLOC (215750) and AFSiD (214989) and the Dutch Fundamenteel Onderzoek
der Materie FOM. We thank N. Collaert and S. Biesemans at IMEC, Leuven for the fabrication of the dopant device.

\bibliography{donors-in-Si}
\end{document}